\documentclass[twocolumn]{aastex61}
\usepackage{amsmath,color,textcomp,url,graphicx,subfigure}
\newcommand{\kms}{\hbox{km\,s$^{-1}$}}

\newcommand{\Mjup}{$M_{\mathrm{Jup}}$}
\newcommand{\Rjup}{$R_{\mathrm{Jup}}$}
\newcommand{\Lbol}{$\log{L_*/L_\odot}$}
\newcommand{\masyr}{$\mathrm{mas}\,\mathrm{yr}^{-1}$}
\newcommand{\teff}{$T_{\rm eff}$}

\newcommand{\simp}{SIMP0136+0933}

\received{2017 April 22}
\revised{2017 May 2}
\accepted{2017 May 3}
\submitjournal{ApJ Letters.}

\shorttitle{SIMP0136+0933 IS LIKELY A MEMBER OF CARINA-NEAR}
\shortauthors{Gagn\'e et al.}

\begin{document}

\title{SIMP~J013656.5+093347 IS LIKELY A PLANETARY-MASS OBJECT IN THE CARINA-NEAR MOVING GROUP}
\author[0000-0002-2592-9612]{Jonathan Gagn\'e}
\affiliation{Carnegie Institution of Washington DTM, 5241 Broad Branch Road NW, Washington, DC~20015, USA}
\affiliation{NASA Sagan Fellow}
\email{jgagne@carnegiescience.edu}
\author[0000-0001-6251-0573]{Jacqueline K. Faherty}
\affiliation{Department of Astrophysics, American Museum of Natural History, Central Park West at 79th St., New York, NY 10024, USA}
\author[0000-0002-6523-9536]{Adam J. Burgasser}
\affil{Center for Astrophysics and Space Sciences, 9500 Gilman Dr., Mail Code 0424, La Jolla, CA 92093, USA}
\author[0000-0003-3506-5667]{\'Etienne Artigau}
\affil{Institute for Research on Exoplanets, Universit\'e de Montr\'eal, D\'epartement de Physique, C.P.~6128 Succ. Centre-ville, Montr\'eal, QC H3C~3J7, Canada}
\author[0000-0002-7904-5484]{Sandie Bouchard}
\affil{Institute for Research on Exoplanets, Universit\'e de Montr\'eal, D\'epartement de Physique, C.P.~6128 Succ. Centre-ville, Montr\'eal, QC H3C~3J7, Canada}
\author[0000-0002-6780-4252]{Lo\"{i}c Albert}
\affil{Institute for Research on Exoplanets, Universit\'e de Montr\'eal, D\'epartement de Physique, C.P.~6128 Succ. Centre-ville, Montr\'eal, QC H3C~3J7, Canada}
\author[0000-0002-6780-4252]{David Lafreni\` ere}
\affil{Institute for Research on Exoplanets, Universit\'e de Montr\'eal, D\'epartement de Physique, C.P.~6128 Succ. Centre-ville, Montr\'eal, QC H3C~3J7, Canada}
\author[0000-0001-5485-4675]{Ren\'e Doyon}
\affil{Institute for Research on Exoplanets, Universit\'e de Montr\'eal, D\'epartement de Physique, C.P.~6128 Succ. Centre-ville, Montr\'eal, QC H3C~3J7, Canada}
\author[0000-0001-8170-7072]{Daniella C. Bardalez Gagliuffi}
\affil{Center for Astrophysics and Space Sciences, University of California, San Diego, 9500 Gilman Dr., Mail Code 0424, La Jolla, CA~92093, USA}

\begin{abstract}
We report the discovery that the nearby ($\sim$\,6\,pc) photometrically variable T2.5 dwarf SIMP~J013656.5+093347 is a likely member of the $\sim$\,200\,Myr-old Carina-Near moving group with a probability of $>$\,99.9\% based on its full kinematics. Our $v\sin i$ measurement of $50.9 \pm 0.8$\,\kms\ combined with the known rotation period inferred from variability measurements provide a lower limit of $1.01 \pm 0.02$\,\Rjup\ on the radius of SIMP0136+0933, an independent verification that it must be younger than $\sim$\,950\,Myr according to evolution models. We estimate a field interloper probability of 0.2\% based on the density of field T0--T5 dwarfs. At the age of Carina-Near, \simp\ has an estimated mass of $12.7 \pm 1.0$\,\Mjup\ and is predicted to have burned roughly half of its original deuterium. SIMP0136+0933 is the closest known young moving group member to the Sun, and is one of only a few known young T dwarfs, making it an important benchmark for understanding the atmospheres of young planetary-mass objects.
\end{abstract}

\keywords{stars: individual (SIMP~J013656.5+093347) --- brown dwarfs --- stars: kinematics and dynamics --- planets and satellites: atmospheres}

\section{INTRODUCTION}\label{sec:intro}

Young brown dwarfs near to and below the deuterium burning mass boundary have the potential to serve as benchmarks in understanding the atmospheres and fundamental properties of gas giant exoplanets, as they share similar temperatures, surface gravities and masses \citep{2015ApJS..219...33G,2016ApJS..225...10F}.

As brown dwarfs cool down with time (e.g. \citealt{2008ApJ...689.1327S}), their masses cannot be determined from effective temperatures only, and their ages must also be constrained. One of the few methods to precisely constrain the ages of brown dwarfs is to identify those that are members of young stellar associations (e.g. see \citealp{2004ARAA..42..685Z,2015IAUS..314...21M}). Recent efforts have been made to identify such objects at the very-low mass end of the brown dwarf regime, using near-infrared large-area surveys (e.g., \citealp{2015ApJS..219...33G,2016ApJ...821..120A,2017arXiv170303774S}).

Since objects in the planetary-mass regime are inherently faint, only about a dozen have been discovered yet, most of which still require confirmation from a parallax or radial velocity measurement (e.g., \citealp{2013ApJ...777L..20L,2015ApJ...808L..20G,2016ApJ...822L...1S,2016ApJ...821L..15K}).

The Carina-Near moving group was discovered and characterized by \cite{2006ApJ...649L.115Z}. It includes a spatially packed core of eight members and a stream of ten additional probable members more loosely distributed in $XYZ$ Galactic coordinates. Based on a comparison of lithium abundance and X-ray luminosity of the Carina-Near members to those of other associations, \cite{2006ApJ...649L.115Z} determined an age of $200 \pm 50$\,Myr for the group.

\cite{2015IAUS..314...21M} notes that no B- or A-type members of Carina-Near are present in Hipparcos \citep{2007AA...474..653V}, and no systematic survey has been published since \cite{2006ApJ...649L.115Z} to identify additional low-mass members. The latest-type member of Carina-Near listed by \cite{2006ApJ...649L.115Z} is the M2.5 dwarf GJ~140~C.

In this paper, we report that the variable T2.5 dwarf \simp\ (SIMP~J013656.5+093347; \citealp{2006ApJ...651L..57A,2009ApJ...701.1534A}) is a likely member of Carina-Near. The BASS-Ultracool survey that led to this discovery is summarized in Section~\ref{sec:buc}. New spectroscopic observations are described in Section~\ref{sec:obs}, and the full 6-dimensional kinematics of \simp\ are discussed in Section~\ref{sec:kin}. In Section \ref{sec:interloper}, it is demonstrated that \simp\ is unlikely a random interloper from the field, and its physical properties are estimated in Section~\ref{sec:props}. Section~\ref{sec:var} discusses the photometric variability of \simp\ in light of its young age. This work is concluded in Section~\ref{sec:conclusion}.

\section{THE BASS-ULTRACOOL SURVEY}\label{sec:buc}

The BANYAN~All-Sky Survey-Ultracool (BASS-Ultracool; J.~Gagn\'e et al., in preparation) was initiated to locate the late-L to T-type members of young moving groups, with the aim to explore the fundamental properties of isolated planetary-mass objects with cold atmospheres (\teff\ $\leqslant$ 1\,500\,K). It relies on a cross-match of large-scale red and near-infrared catalogs such as 2MASS \citep{2006AJ....131.1163S}, AllWISE \citep{2010AJ....140.1868W}, and Pan-STARRS1 \citep{2016arXiv161205560C} to identify high proper motion objects with red $W1-W2$ AllWISE colors, for which moving group membership is assessed with the Bayesian Analysis for Nearby Young AssociatioNs~II tool (BANYAN~II; \citealt{2014ApJ...783..121G}), and its successor BANYAN~$\Sigma$ (J. Gagn\'e et al., in preparation). First discoveries from this survey include the T5.5 dwarf SDSS~J111010.01+011613.1 as a $\sim$\,10--12\,\Mjup\ member of AB~Doradus   \citep{2015ApJ...808L..20G}; 2MASS~J09553336--0208403, a young and unusually red L7-type interloper to the TW~Hya association \citep{2017ApJS..228...18G}; and an $\sim$\,L7+T5 spectral binary candidate member of AB~Doradus (D.~Bardalez-Gagliuffi et al., submitted to ApJL).

The \cite{2006ApJ...649L.115Z} list of members and probable members of Carina-Near were cross-matched with the {\it Gaia} data release 1 \citep{2016AA...595A...4L}, which yielded 12 parallax and proper motion measurements that are more precise than those previously available. New radial velocity measurements from \cite{2015AA...573A.126D} were also included for six members. The resulting properties of Carina-Near bona fide members are listed in Table~\ref{tab:carn}.

%Table of Carina-Near members
\begin{deluxetable*}{lcclccccc}
\renewcommand\arraystretch{0.9}
\tabletypesize{\small}
\tablecaption{Updated kinematics of Carina-Near bona fide members. \label{tab:carn}}
\tablehead{\colhead{Name} & \colhead{R.A.} & \colhead{Decl.} & \colhead{Spectral} & \colhead{$\mu_\alpha\cos\delta$} & \colhead{$\mu_\delta$} & \colhead{Rad. velocity} & \colhead{Parallax} & \colhead{Ref.\tablenotemark{a}}\\
\colhead{} & \colhead{(hh:mm:ss.ss)} & \colhead{(dd:mm:ss.s)} & \colhead{Type} & \colhead{(\masyr)} & \colhead{(\masyr)} & \colhead{(\kms)} & \colhead{(mas)} & \colhead{}}
\vspace{-0.2cm}\startdata
\sidehead{\textbf{Core Members}\vspace{-0.1cm}}
HD~59704 & 07:29:31.38 & --38:07:20.6 & F7 & $-27.32 \pm 0.03$ & $68.06 \pm 0.04$ & $24.7 \pm 1.3$ & $19.24 \pm 0.27$ & 1,2,1,3,1\\
HD~62850 & 07:42:35.96 & --59:17:48.4 & G2/3 & $-53.93 \pm 0.07$ & $158.63 \pm 0.06$ & $17.1 \pm 0.4$ & $30.73 \pm 0.22$ & 1,2,1,3,1\\
HD~62848 & 07:43:21.40 & --52:09:48.5 & G0 & $-56.89 \pm 0.03$ & $157.40 \pm 0.03$ & $20.5 \pm 0.5$ & $29.16 \pm 0.56$ & 1,2,1,4,1\\
HD~63581 & 07:46:14.70 & --59:48:48.4 & K0~IV/V & $-57.3 \pm 0.1$ & $154.6 \pm 0.1$ & $18.1 \pm 0.1$ & $30.39 \pm 0.55$ & 1,5,1,3,1\\
HD~63608 & 07:46:16.86 & --59:48:31.9 & K0 & $-52.3 \pm 0.4$ & $153.2 \pm 0.4$ & $16.9 \pm 0.3$ & $29.76 \pm 0.23$ & 1,2,1,3,1\\
HR~3070 & 07:49:12.90 & --60:17:01.3 & F1 & $-37.9 \pm 0.3$ & $140.3 \pm 0.3$ & $16.9 \pm 0.4$ & $28.9 \pm 0.3$ & 6,2,6,4,6\\
CPD--52~1153~A & 07:20:21.36 & --52:18:39.3 & F2 & $-36.88 \pm 0.05$ & $146.69 \pm 0.05$ & $17.3 \pm 1.5$ & $32.75 \pm 0.69$ & 1,2,1,7,1\\
CPD--52~1153~B & 07:20:21.81 & --52:18:31.2 & G0 & $\cdots$ & $\cdots$ & $15.5 \pm 3.9$ & $\cdots$ & 1,2,,7,\\
\sidehead{\textbf{Stream Members}\vspace{-0.1cm}}
GJ~140~AB & 03:24:06.65 & +23:47:06.7 & M1+M1 & $225.3 \pm 4.5$ & $-131.4 \pm 3.7$ & $19 \pm 2$ & $51.3 \pm 4.7$ & 1,2,6,2,6\\
GJ~140~C & 03:24:13.08 & +23:46:17.4 & M2.5 & $199 \pm 8$ & $-112 \pm 8$ & $18 \pm 3$ & $\cdots$ & 1,2,8,2,\\
GJ~358 & 09:39:45.68 & --41:03:57.9 & M2 & $-526.6 \pm 1.4$ & $356.4 \pm 1.4$ & $18 \pm 3$ & $105.6 \pm 1.6$ & 1,2,6,2,6\\
HD~108574 & 12:28:04.19 & +44:47:39.4 & F7 & $-181.8 \pm 0.1$ & $-4.5 \pm 0.1$ & $-2.3 \pm 0.3$ & $22.01 \pm 0.24$ & 1,2,1,4,1\\
HD~108575 & 12:28:04.54 & +44:47:30.5 & G0 & $-180.7 \pm 0.1$ & $0.5 \pm 0.1$ & $-1.3 \pm 0.9$ & $21.84 \pm 0.24$ & 1,2,1,4,1\\
HD~103742 & 11:56:42.11 & --32:16:05.5 & G3 & $-172.0 \pm 0.4$ & $-8.3 \pm 0.3$ & $5.9 \pm 0.1$ & $27.97 \pm 0.33$ & 1,2,1,3,1\\
HD~103743 & 11:56:43.57 & --32:16:02.8 & G3 & $-178.8 \pm 0.6$ & $-7.1 \pm 0.4$ & $6.9 \pm 0.5$ & $28.23 \pm 0.26$ & 1,2,1,3,1\\
GJ~900 & 23:35:00.62 & +01:36:19.9 & M0 & $340.9 \pm 0.1$ & $27.14 \pm 0.08$ & $-9.4 \pm 1.1$ & $48.17 \pm 0.31$ & 1,2,1,4,1\\
GJ~907.1 & 23:48:25.93 & --12:59:14.6 & K8 & $230.2 \pm 3.3$ & $21.0 \pm 2.0$ & $-8.4 \pm 0.5$ & $35.46 \pm 2.2$ & 1,2,6,9,6\\
\enddata
\tablenotetext{a}{All members were determined by \cite{2006ApJ...649L.115Z}. References for positions, spectral types, proper motions, radial velocities and parallaxes are given in this respective order for each individual object.}
\tablerefs{(1)~\citealt{2016AA...595A...4L}, (2)~\citealt{2006ApJ...649L.115Z}, (3)~\citealt{2015AA...573A.126D}, (4)~\citealt{2006AstL...32..759G}, (5)~\citealt{2006AJ....132..161G}, (6)~\citealt{2007AA...474..653V}, (7)~\citealt{2004AA...418..989N}, (8)~\citealt{2013AJ....145...44Z}, (9)~\citealt{2006AA...460..695T}.}
\end{deluxetable*}

A multivariate Gaussian model was fit to the resulting kinematics:
\begin{align}
	\mathcal{P}\left(\bar x|\bar x_0, \bar{\bar\Sigma}\right) &= \frac{\exp{\left(-\left(\bar{x}-\bar{x}_0\right)^T\bar{\bar{\Sigma}}^{-1}\left(\bar{x}-\bar{x}_0\right)/2\right)}}{\sqrt{\left(2\pi\right)^6\left|\bar{\bar\Sigma}\right|}},
\end{align}
where $\bar x$ is a 6-dimensional vector of coordinates in $XYZ$ Galactic coordinates (in pc) and $UVW$ space velocities (in \kms) in a right-handed system where $U$ points toward the Galactic center, $\bar x_0$ is a 6-dimensional vector representing the center of the group, $\bar{\bar\Sigma}$ is the covariance matrix of all members, and $\left|\bar{\bar\Sigma}\right|$ is the determinant of the covariance matrix.

The diagonal elements of $\bar{\bar\Sigma}$ represent variances in the $XYZUVW$ directions, while the off-diagonal elements represent correlations between spatial and kinematic coordinates, which can be related to rotation angles in 6-dimensional space. This multivariate model is a generalization of the freely rotating 3D ellipsoid models used in BANYAN~II, as it allows for correlations in the spatial and kinematic coordinates. The resulting model parameters are (in order of $XYZUVW$):
\begin{align}
	\bar x_0 &=  \begin{bmatrix}
    -2.39 & -18.55 & -4.21 & -26.49 & -17.52 & -2.34
    \end{bmatrix},\notag\\
    \bar{\bar\Sigma} &= \begin{bmatrix}
    	60.5 & -16.7 & -15.0 & 3.8 & 2.0 & 6.4\\
		-16.7 & 434.3 & 47.3 & -46.4 & 7.9 & -1.6\\
		-15.0 & 47.3 & 300.1 & -15.7 & -23.1 & -23.4\\
		3.8 & -46.4 & -15.7 & 10.2 & 0.5 & -0.5\\
		2.0 & 7.9 & -23.1 & 0.5 & 3.5 & 2.2\\
		6.4 & -1.6 & -23.4 & -0.5 & 2.2 & 3.8
    \end{bmatrix},
\end{align}
in units of pc and \kms.
   
In this survey, \simp\ was identified as a candidate member of Carina-Near from a cross-match of 2MASS and AllWISE sources with colors redder than $W1-W2 = 0.6$, which corresponds to spectral types later than $\sim$\,L8 (field dwarfs) or $\sim$\,L5 (young dwarfs; \citealt{2016ApJS..225...10F}). This T2.5 dwarf was discovered as part of the SIMP survey \citep{2006ApJ...651L..57A,2016ApJ...830..144R} and has been the subject of extensive photometric follow-up due to its photometric variability (e.g., see \citealp{2009ApJ...701.1534A,2013AN....334...40M,2014ApJ...793...75R,2016arXiv160903587C}).

A proper motion derived from 2MASS and AllWISE alone categorized it as a high-probability ($>$\,99.9\%) candidate member of the Carina-Near moving group using a preliminary version of the BANYAN~$\Sigma$ tool, with a kinematic distance ($5.6\pm 0.3$\,pc) placing it along the sequence of known T dwarfs in near-infrared color-magnitude diagrams. Using the updated astrometry from \cite{2016AJ....152...24W} preserved a high Bayesian membership probability of $>$\,99.9\%, with a statistical radial velocity prediction of $9.4 \pm 0.8$\,\kms\ if it is a member of Carina-Near.

\section{OBSERVATIONS}\label{sec:obs}

\simp\ was observed with the Near InfraRed Spectrometer (NIRSPEC; \citealt{2000SPIE.4008.1048M}) on the Keck~II Telescope on 2013 October 16 and 2016 February 2. The first night had moderate cloud coverage with $\sim$60\% humidity and a 1$\arcsec$ seeing, while the second night was clear with a 0$\farcs$8 seeing. The high-dispersion cross-dispersed mode with the NIRSPEC-7 filter and the 0$\farcs$432-wide slit were used, yielding a resolving power of $\lambda/\Delta\lambda \approx$\,20,000 over 2.00--2.39\,\micron. Two exposures of 900\,s (2013 October) and 750\,s (2016 February) were obtained at airmasses of 1.11 and 1.15, yielding signal-to-noise ratios of $\sim$20 and $\sim$13 per pixel after reduction. The A0V-type standard HD~6457 was observed immediately after \simp\ at a similar airmass for telluric correction. A single NeArXeKr calibration lamp exposure and ten 4.4s ``on'' and ``off'' flat-field exposures were obtained at the end of each night for wavelength calibration and correction of pixel-to-pixel variations in detector response. The data were reduced with a modified version of the REDSPEC package as described in \cite{2015AJ....149..104B}.  

\begin{figure*}[p]
	\centering
	\subfigure{\includegraphics[width=0.48\textwidth]{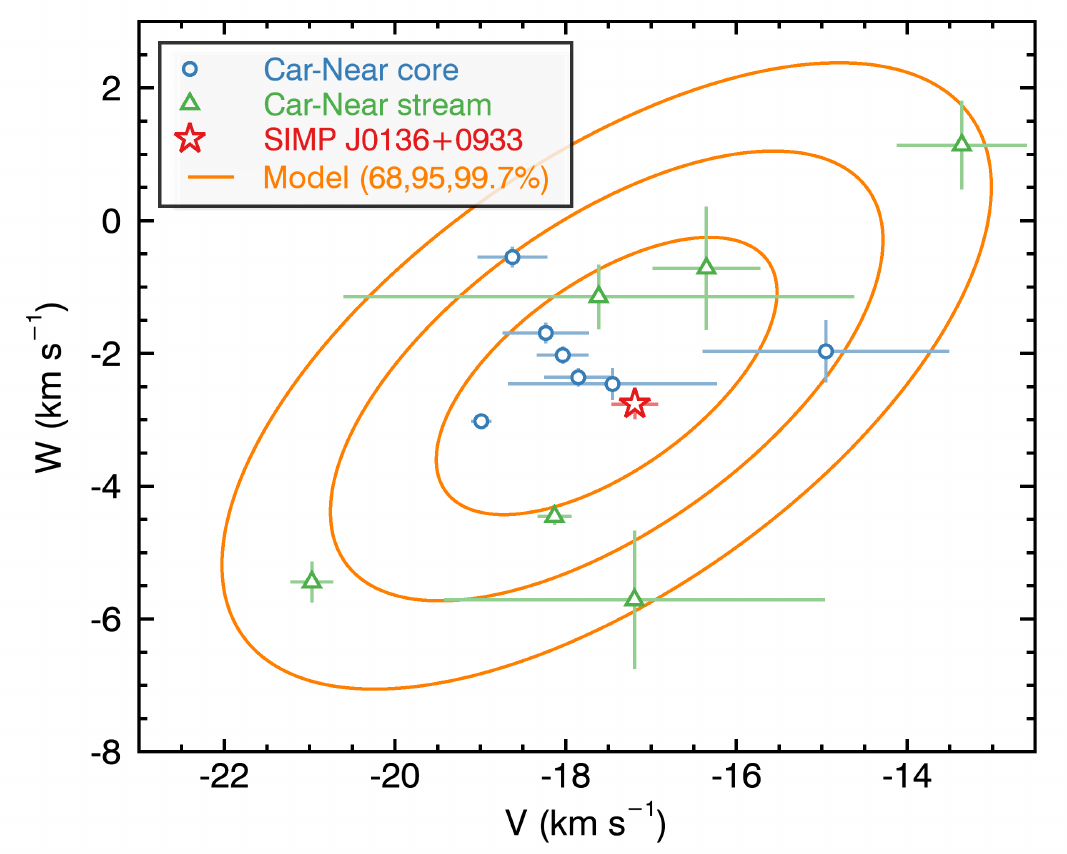}\label{fig:VW}}
    \subfigure{\includegraphics[width=0.48\textwidth]{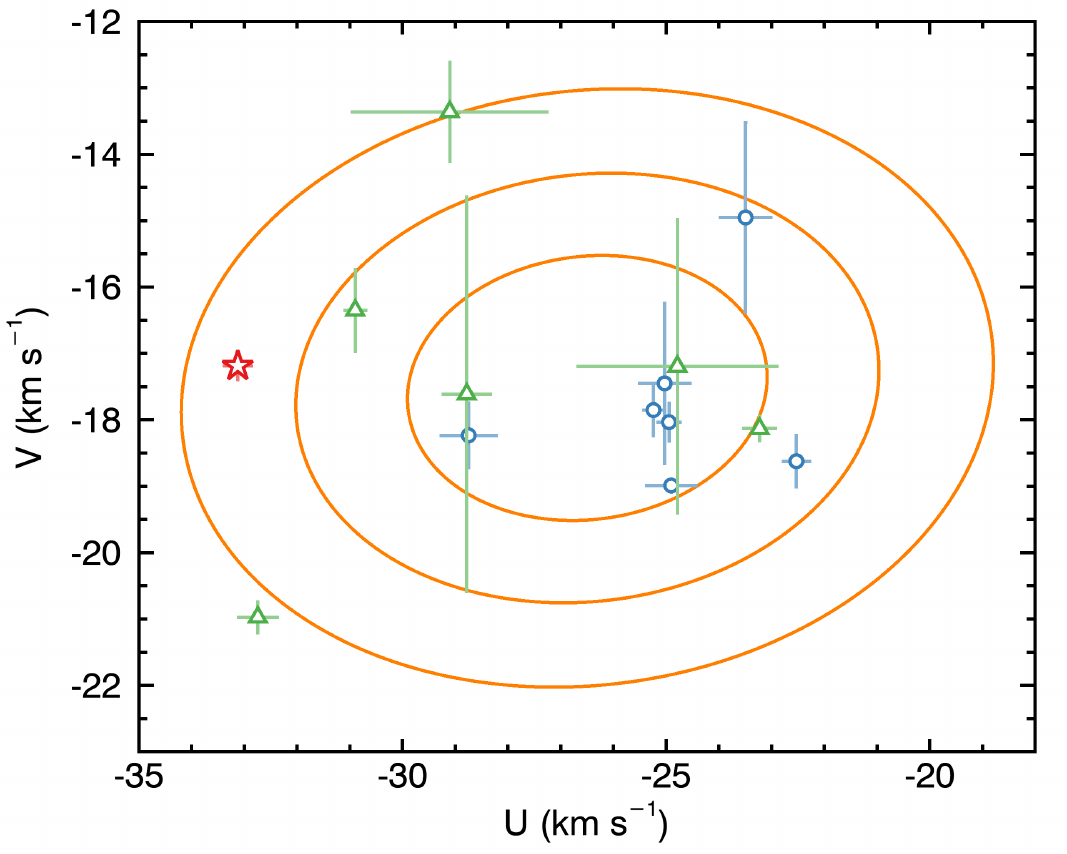}\label{fig:UV}}
    \subfigure{\includegraphics[width=0.48\textwidth]{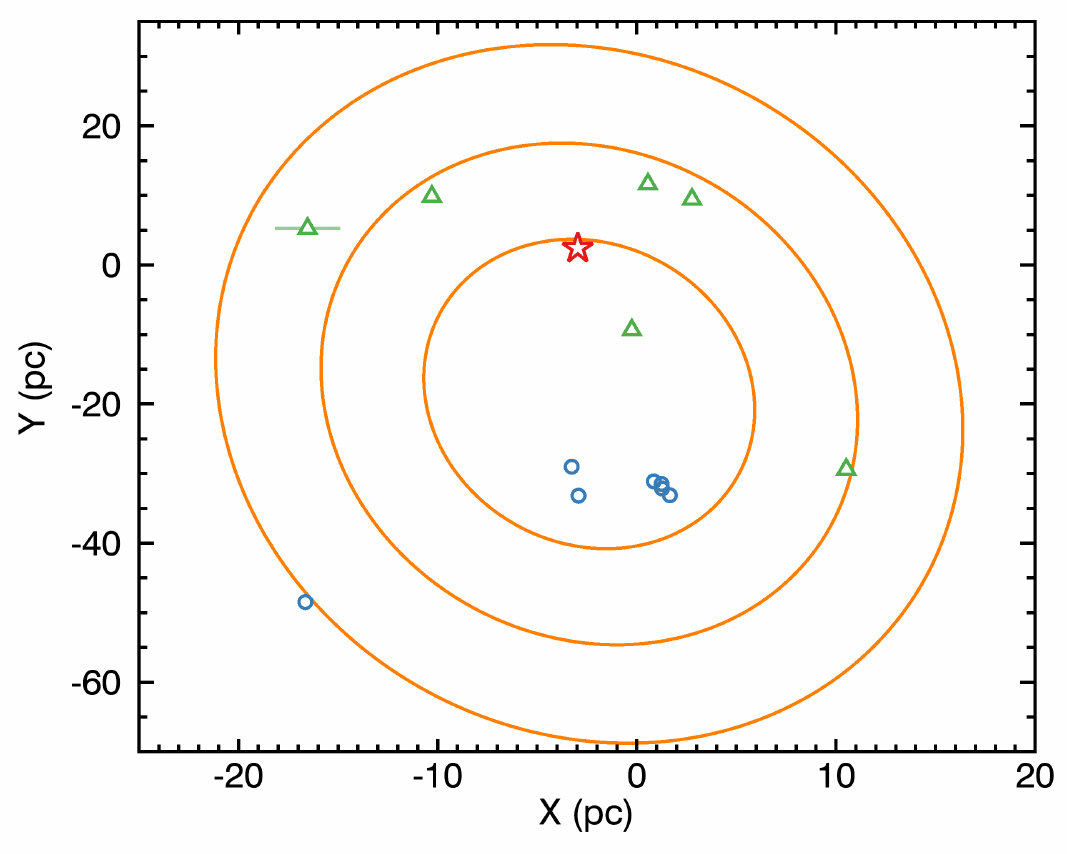}\label{fig:XY}}
    \subfigure{\includegraphics[width=0.48\textwidth]{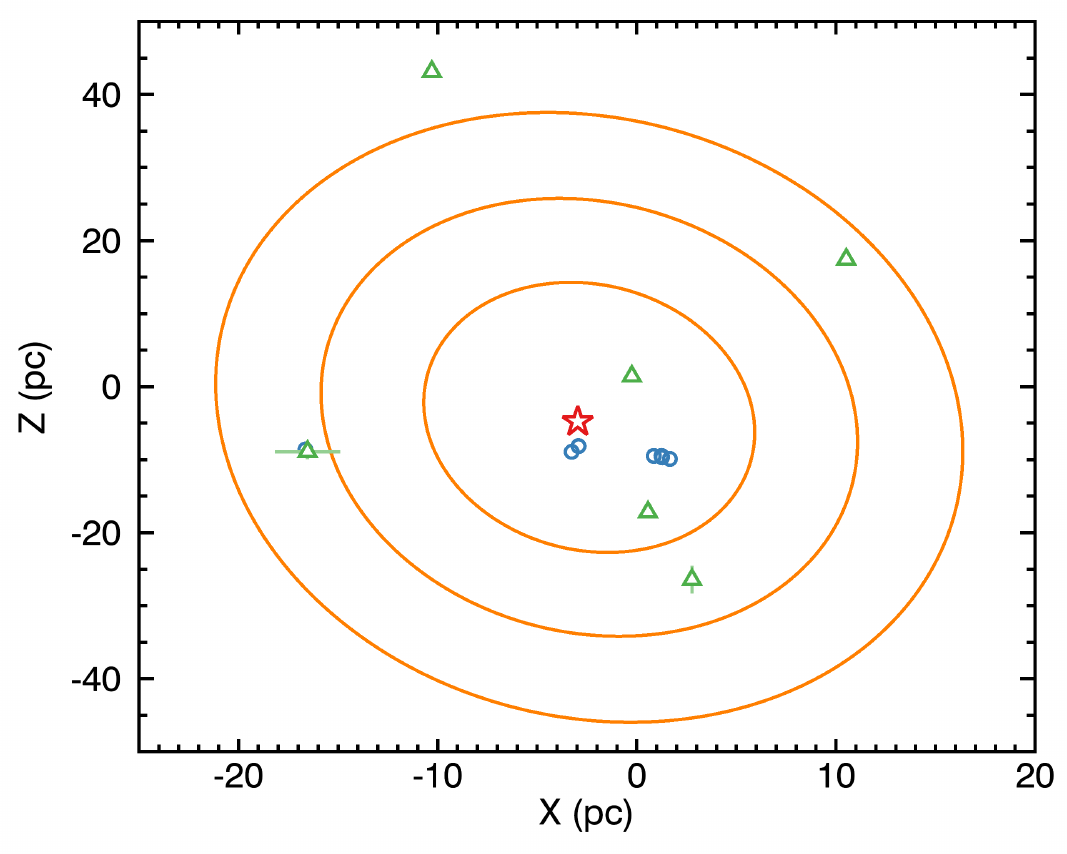}\label{fig:XZ}}
	\caption{Galactic coordinates $XYZ$ and space velocities $UVW$ of \simp\ (red star) compared to bona fide members of the Carina-Near core (blue circles) and stream (green triangles) as defined by \cite{2006ApJ...649L.115Z}, and the multivariate Gaussian model of Carina-Near used in BANYAN~$\Sigma$ (orange lines of 68\%, 95\% and 99.7\% confidence).}
	\label{fig:kinematics}
\end{figure*}

\begin{figure}[!htbp]
	\centering
	\includegraphics[width=0.48\textwidth]{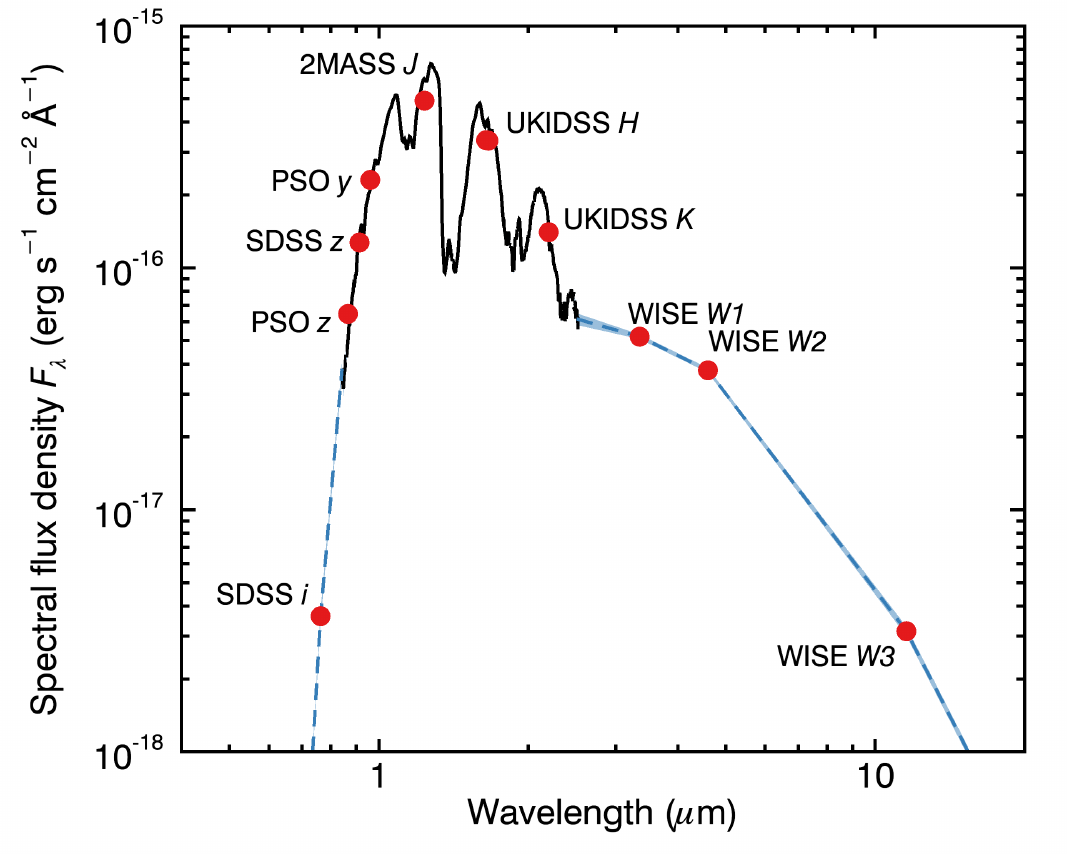}
	\caption{Spectral energy distribution of \simp, built from the GNIRS spectrum of \cite{2016ApJ...830..144R} and optical to infrared photometry listed in Table~\ref{tab:properties}, with the method of \cite{2015ApJ...810..158F}.}
	\label{fig:sed}
\end{figure}

%Table of properties
\begin{deluxetable}{lcc}
\renewcommand\arraystretch{0.9}
\tabletypesize{\small}
\tablecaption{Properties of SIMP~J013656.5+093347 \label{tab:properties}}
\tablehead{\colhead{Property} & \colhead{Value} & \colhead{Reference}}
\vspace{-0.2cm}\startdata
\sidehead{\textbf{Position and Kinematics}\vspace{-0.1cm}}
R.A. & 01:36:56.62 & 1\\
Decl.  & +09:33:47.3 & 1\\
$\mu_\alpha\cos\delta$ (\masyr) & $1222.70 \pm 0.78$ & 1\\
$\mu_\delta$ (\masyr) & $0.5 \pm 1.2$ & 1\\
RV (\kms) & $11.5 \pm 0.4$ & 2\\
Trigonometric distance (pc) & $6.139 \pm 0.037$ & 1\\
$X$ (pc) & $-2.967 \pm 0.018$ & 2\\
$Y$ (pc) & $2.384 \pm 0.014$ & 2\\ 
$Z$ (pc) & $-4.817 \pm 0.029$ & 2\\
$U$ (\kms) & $-33.12 \pm 0.26$ & 2\\
$V$ (\kms) & $-17.19 \pm 0.20$ & 2\\
$W$ (\kms) & $-2.76 \pm 0.32$ & 2\vspace{-0.2cm}\\
\sidehead{\textbf{Photometric Properties}\vspace{-0.1cm}}
$i_{\rm AB}$ (\emph{SDSS12}) & $20.79 \pm 0.06$ & 3\\
$z_{\rm AB}$ (\emph{Pan-STARRS1}) & $17.33\pm0.02$ & 4\tablenotemark{a}\\
$z_{\rm AB}$ (\emph{SDSS12}) & $16.55 \pm 0.02$ & 3\\
$y_{\rm AB}$ (\emph{Pan-STARRS1}) & $15.72\pm0.03$ & 4\tablenotemark{a}\\
$J$ (\emph{2MASS}) & $13.46 \pm 0.03$ & 5\\
$H$ (\emph{UKIDSS}) & $12.809 \pm 0.002$ & 6\\
$K$ (\emph{UKIDSS})& $12.585 \pm 0.002$ & 6\\
$W1$ (\emph{AllWISE}) & $11.94 \pm 0.02$ & 7\\
$W2$ (\emph{AllWISE}) & $10.96 \pm 0.02$ & 7\\
$W3$ (\emph{AllWISE}) & $9.74 \pm 0.05$ & 7\vspace{-0.2cm}\\
\sidehead{\textbf{Spectroscopic Properties}\vspace{-0.1cm}}
Spectral type & T2.5$ \pm 0.5$ & 8\\
\ion{K}{1} at 1.169\,\micron\ (\AA) & $8.6 \pm 1.0$ & 2\\
\ion{K}{1} at 1.177\,\micron\ (\AA) & $11.7 \pm 1.1$ & 2\\
\ion{K}{1} at 1.243\,\micron\ (\AA) & $5.4 \pm 0.3$ & 2\\
\ion{K}{1} at 1.254\,\micron\ (\AA) & $8.5 \pm 0.3$ & 2\vspace{-0.2cm}\\
\sidehead{\textbf{Fundamental Properties}\vspace{-0.1cm}}
Age (Myr) & $200 \pm 50$ & 9\\
Mass (\Mjup) & $12.7 \pm 1.0$ & 2\\
Radius (\Rjup) & $1.22 \pm 0.01$ & 2\\
\teff\ (K) & $1098 \pm 6$ & 2\\
$\log g$ & $4.31 \pm 0.03$ & 2\\
\Lbol & $-4.688 \pm 0.005$ & 2\\
Rotation period (h) & $2.425 \pm 0.003$ & 10\tablenotemark{b}\\
$v\sin i$ (\kms) & $50.9 \pm 0.8$ & 2\\
$i$ (\textdegree) & $55.9_{-1.5}^{+1.6}$ & 2\\
\enddata
\tablenotetext{a}{Average and standard deviation from five PS1 measurements. The dispersion is larger than quoted uncertainties, likely indicating that variability is also present at red-optical wavelengths.}
\tablenotetext{b}{S. Bouchard et al. (in prep.) used an improved method and $\sim$\,8\,hr of $J$-band monitoring over 2 nights to improve the previous measurement of $2.3895 \pm 0.0005$ \citep{2009ApJ...701.1534A}.}
\tablerefs{(1)~\citealt{2016AJ....152...24W}, (2)~This work, (3)~\citealt{2015ApJS..219...12A}, (4)~\citealt{2016arXiv161205560C}, (5)~\citealt{2006AJ....131.1163S}, (6)~\citealt{2007MNRAS.379.1599L}, (7)~\citealt{2014ApJ...783..122K}, (8)~\citealt{2006ApJ...651L..57A}, (9)~\citealt{2006ApJ...649L.115Z},
(10)~S. Bouchard et al., in preparation.}
\end{deluxetable}

\section{THE KINEMATICS OF SIMP0136+0933}\label{sec:kin}

Spectral data in order 33 (2.29--2.33\,\micron) were forward-modeled using the Markov Chain Monte Carlo analysis described in detail in \cite{2016ApJ...827...25B}. We used the atmosphere models of \cite{2012RSPTA.370.2765A} and the telluric transmission spectrum from the Solar atlas of \cite{1991aass.book.....L}. Radial and rotational velocities and their uncertainties were determined by marginalizing over the Markov chains, yielding consistent results between the two epochs (differences of 0.4 and 0.1\,\kms, respectively). Combining these values as an error-weighted average, we determine a radial velocity of $11.5 \pm 0.4$\,\kms\ and $v\sin i = 50.9 \pm 0.8$\,\kms\ for \simp.

The $XYZ$ Galactic coordinates and $UVW$ space velocities of \simp\ were calculated by adding the radial velocity measurement to the set of kinematic measurements available in the literature in a $10^4$-elements Monte Carlo simulation that assumes Gaussian measurement errors, and are reported in Table~\ref{tab:properties}. These coordinates place \simp\ at $20.9 \pm 11.6$\,pc and $6.4 \pm 1.6$\,\kms\ from the core of the Carina-Near multivariate Gaussian model (the largest kinematic deviation occurs in the $U$ direction, see Figure~\ref{fig:kinematics}). This makes \simp\ a likely member of the Carina-Near stream, which is more spatially extended than its core. The second-nearest Carina-Near member to the Sun, the M2 dwarf GJ~358 ($9.5 \pm 0.1$\,pc; \citealt{2007AA...474..653V}), is also a member of its stream \citep{2006ApJ...649L.115Z}. At a distance of $\sim$\,6.1\,pc, \simp\ is the nearest known member of any young moving groups, and among the nearest young objects known to date, along with the M4+M5 EQ~Peg~AB system at $6.18 \pm 0.06$\,pc \citep{2007AA...474..653V,2013ApJ...778....5Z}, the M5 dwarf AP~Col at $8.39 \pm 0.07$\,pc \citep{2011AJ....142..104R}, and a few possibly young X-ray active mid-M dwarfs from the sample of \citeauthor{2006AJ....132..866R} (\citeyear{2006AJ....132..866R}; see their Table~4). \simp\ is also among the 100 nearest systems to the Sun.

\section{FIELD INTERLOPER PROBABILITY}\label{sec:interloper}

The probability that \simp\ is a random field interloper was estimated with the field density of T0--T5 dwarfs measured by \cite{2010AA...522A.112R} and a synthetic population of $10^7$ objects within $6.14$\,pc, drawn from the $XYZUVW$ distribution of stars in the Galactic neighborhood using the Besan\c con model \citep{2012AA...538A.106R}. The separation between each object and the Carina-Near multivariate Gaussian model in $XYZUVW$ space was then calculated using the Mahalanobis distance \citep{Mahalanobis:1936va}, given by:
\begin{align}
	\mathcal{M} &= \sqrt{\left(\bar{x}-\bar{x}_0\right)^T\bar{\bar{\Sigma}}^{-1}\left(\bar{x}-\bar{x}_0\right)},\label{eqn:mahal}
\end{align}
where $\bar x$, $\bar x_0$ and $\bar{\bar\Sigma}$ are defined in Section~\ref{sec:buc}. The Mahalanobis distance is a generalized N$\sigma$ distance that accounts for the size and orientation of the multivariate Gaussian model in 6-dimensional space.

A total of 14\,609 out of $10^7$ synthetic objects were found to have a Mahalanobis distance at least as small as that of \simp\ ($\mathcal{M} = 3.28$). Adjusting this result to the T0--T5 field density of $2.0_{-0.2}^{+0.3}\cdot 10^{-3}$\,objects\,pc$^{-3}$ measured by \cite{2010AA...522A.112R}, this corresponds to $4.52_{-0.3}^{+0.4} \cdot 10^{-3}$ expected occurrences. Based on Poisson statistics, the detection of at least one early T dwarf such as \simp\ thus has a 0.2\% probability of being a  chance event. This analysis does not assume a young age for \simp.

\begin{figure*}[!htbp]
\centering
\subfigure[Deuterium burning]{\includegraphics[width=0.43\textwidth]{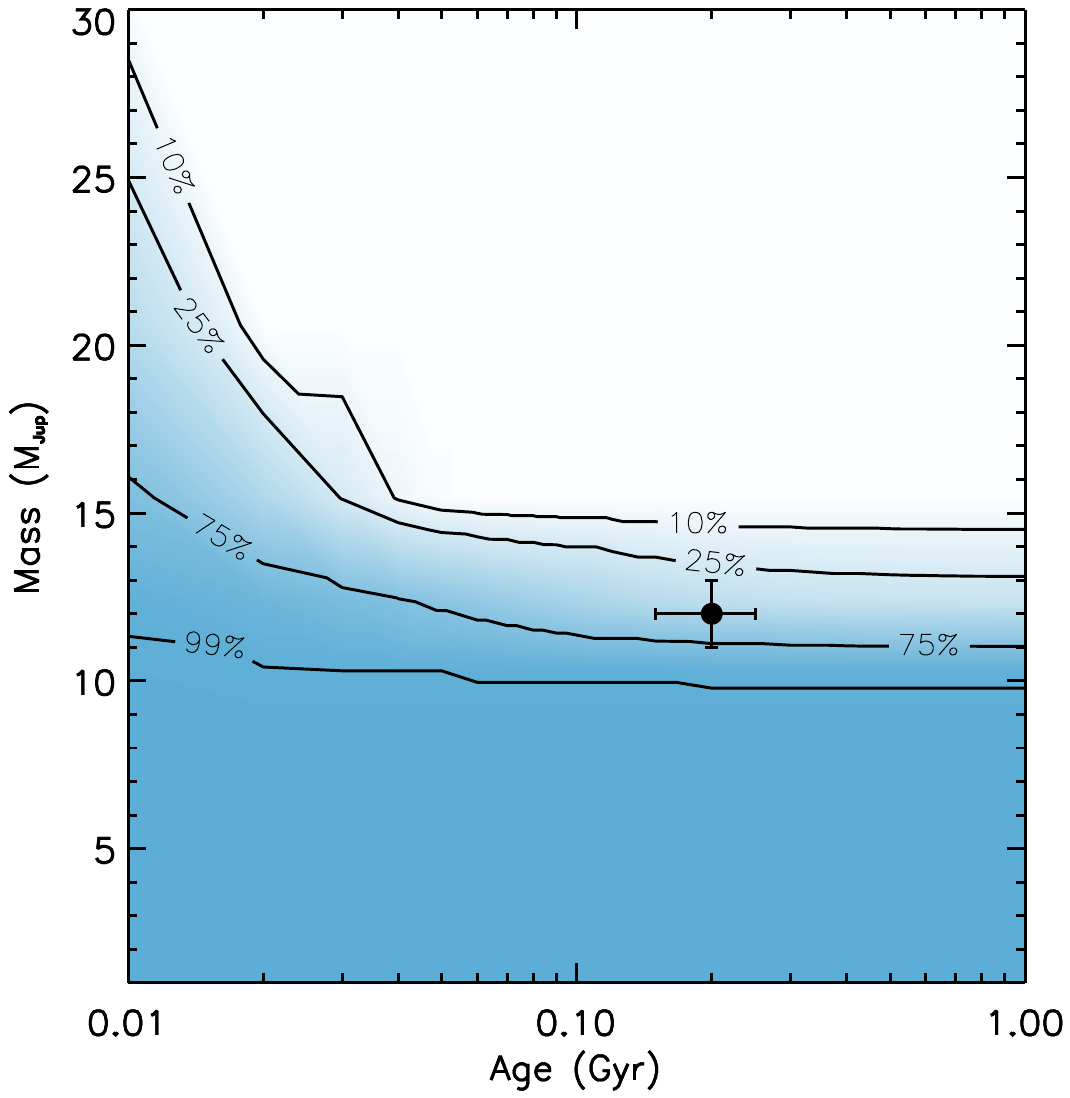}\label{fig:d2}}
\subfigure[Minimum radius]{\includegraphics[width=0.556\textwidth]{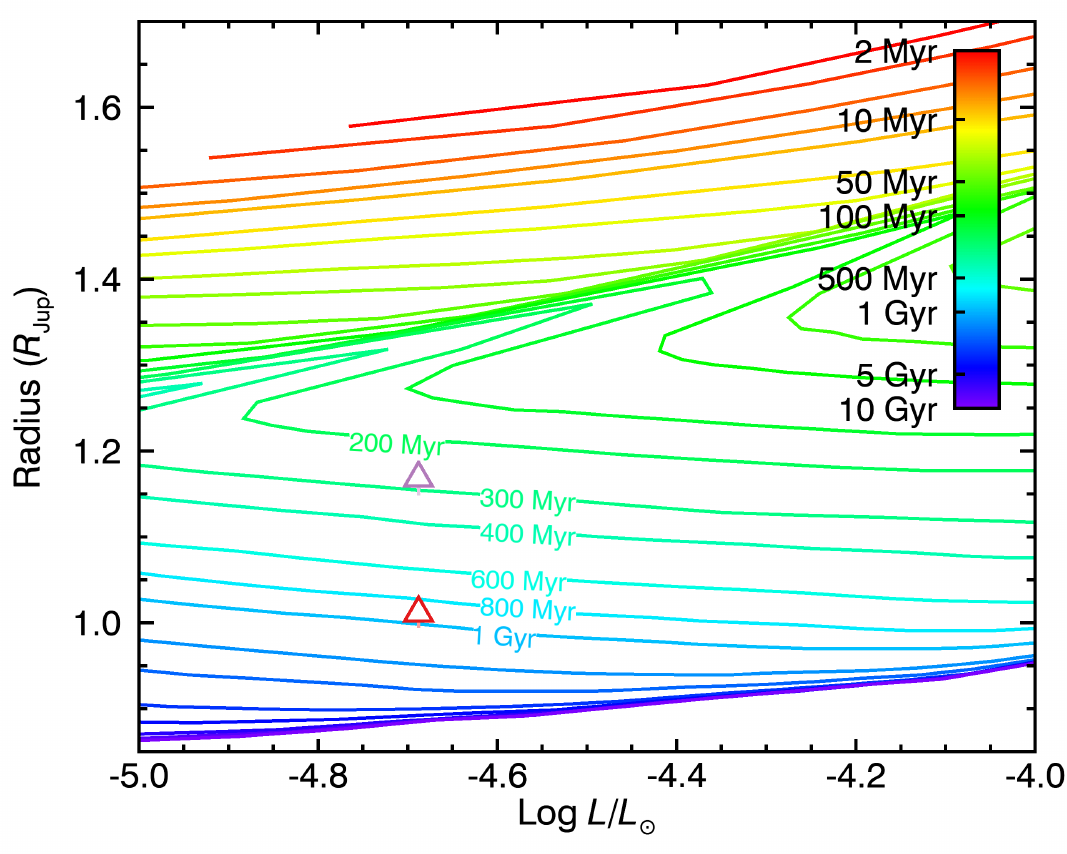}\label{fig:radius}}
\caption{Fundamental properties of \simp. Left panel: deuterium fraction left as a function of age at the brown dwarf/planetary boundary. The models of \cite{2012RSPTA.370.2765A} at the age and estimated mass of \simp\ suggest that it should only have partially depleted its deuterium content. Right panel: minimum radius of \simp\ (red triangle) obtained from $v\sin i$ and rotational periods only; and minimum radius when the $i < 60$\textdegree\ constraint is added (purple triangle); and the models of \citeauthor{2008ApJ...689.1327S} (\citeyear{2008ApJ...689.1327S}; thick colored lines).}
\label{fig:fundamental}
\end{figure*}

\section{FUNDAMENTAL PROPERTIES}\label{sec:props}

The spectral energy distribution of \simp\ was built from a combination of the literature SpeX spectrum \citep{2008ApJ...681..579B} with SDSS, Pan-STARRS1, 2MASS and AllWISE photometry (see Table~\ref{tab:properties}), and is displayed in Figure~\ref{fig:sed}. The method of \cite{2015ApJ...810..158F} was used to perform an empirical measurement of its bolometric luminosity \Lbol $= -4.688 \pm 0.005$, and the \cite{2008ApJ...689.1327S} models at $200 \pm 50$\,Myr were used to infer a radius of $1.22 \pm 0.01$\,\Rjup. These properties were converted to an effective temperature using the Stefan-Boltzmann law, yielding \teff\ = $1098 \pm 6$\,K, at the lower-end of T2--T3 field dwarf temperatures \citep{2015ApJ...810..158F}. This is consistent with the observed trend that young brown dwarfs tend to have slightly lower effective temperatures at a given spectral type \citep{2006ApJ...651.1166M,2012ApJ...752...56F,2015ApJS..219...33G}.

A corresponding mass and surface gravity of $12.7 \pm 1.0$\,\Mjup\ and $\log g = 4.31 \pm 0.03$ were obtained from the \cite{2008ApJ...689.1327S} evolutionary models. As displayed in Figure~\ref{fig:d2}, a comparison of these values with the models of \cite{2012RSPTA.370.2765A} implies that \simp\ should have burned roughly half of its deuterium content. The bolometric luminosity of \simp\ is slightly fainter than average T2.5 dwarfs \citep{2015ApJ...810..158F} although it is located within the 1$\sigma$ range of the field distribution. The $\sim$\,150\,Myr-old T4.5 dwarf GU~Psc~b is similarly fainter than the field dwarfs sequence and yields a comparable \Lbol = $−4.87 \pm 0.10$ \citep{2015ApJ...810..158F}.

The $v\sin i$ measurement obtained in Section~\ref{sec:kin} was combined with the rotation period of \cite{2009ApJ...701.1534A} to constrain the inclination of \simp\ to $i = 55.9_{-1.5}^{+1.6}$\textdegree, assuming the \cite{2008ApJ...689.1327S} radius at the age of Carina-Near. Without making any assumption on the age of \simp, the radius of \simp\ can be constrained to a lower limit of $1.01 \pm 0.02$\,\Rjup, which would correspond to it being observed equator~on, or to larger radii at other inclinations. This radius lower limit can be translated to an upper age limit of $910_{-110}^{+26}$\,Myr, or an upper mass limit of $42.6_{-2.4}^{+2.5}$\,\Mjup, based on the \cite{2008ApJ...689.1327S} models and the measured bolometric luminosity (see Figure~\ref{fig:radius}). Adding the upper limit of this age constraint to the analysis presented in Section~\ref{sec:interloper} further reduces the field interloper probability down to 0.0001\%.

A preliminary analysis of the harmonics in the long-term light curve of \simp\ which assumes that its variability is dominated by a single spot constrains its inclination to $i < 60$\textdegree\ (S.~Bouchard et al., in preparation). If the single-spot hypothesis can be confirmed, it will further constrain the radius to values larger than $1.17 \pm 0.02$\,\Rjup, and a model-dependent age below $280_{-30}^{+40}$\,Myr, which would corroborate the Carina-Near membership and the planetary mass of \simp.

The method of \cite{2003ApJ...596..561M} was used to measure \ion{K}{1} equivalent widths of $8.6 \pm 1.0$, $11.7 \pm 1.1$, $5.4 \pm 0.3$, and $8.5 \pm 0.3$\,\AA\ (at 1.169, 1.177, 1.243, and 1.254\,\micron, respectively) from the GNIRS spectrum. These values are consistent with the field population of T2--T3 dwarfs \citep{2003ApJ...596..561M}, which is not unexpected given that the slightly younger T5.5 dwarf SDSS~J111010.01+011613.1 ($\sim$\,150\,Myr; \citealp{2015ApJ...808L..20G}) has \ion{K}{1} equivalent widths consistent or slightly weaker than those of field T5 dwarfs \citep{2017ApJ...838...73M}.

\section{VARIABILITY AND AGE}\label{sec:var}

Surface gravity is a key parameter in the description of dust behavior in brown dwarf atmospheres. Among L dwarfs, thicker cloud decks are expected among lower gravity objects due to the slower settling rates. Thick cloud decks lead to a redistribution of the near-infrared spectral energy distribution to longer wavelengths, leading to redder infrared colors for low gravity, young objects \citep{2016ApJS..225...10F}. As dust-bearing clouds of varying thicknesses are generally invoked to explain such variability, a gravity-dependence of variability behavior can be expected. \cite{2015ApJ...799..154M} included a sample of low-gravity L~dwarfs in the large SPITZER sample of photometric observations, and found tentative indications that the low-gravity objects displayed larger photometric amplitudes even though the fraction of variable objects appeared to be independent of surface gravity. The detection of large (7-10\%) photometric $J$-band variability in very-low gravity L~dwarfs (PSO~J318.5338--22.8603, WISEP~J004701.06+680352.1; \citealp{2015ApJ...813L..23B,2016ApJ...829L..32L}) appears to corroborate these findings, as earlier surveys at these wavelengths failed to uncover $>$\,4\% variability among L dwarfs.

Figure~\ref{fig:var} compiles $J$, 3.6\,\micron\ and 4.5\,\micron\ variability detections in a spectral-type versus color diagram. The detections to date suggest a higher fraction of high-amplitude variables among very red L-type dwarfs, but some of these discoveries were made from surveys explicitly targeting red, low-gravity objects. Overall, this tentative correlation between surface gravity and variability amplitude has yet to be set on firm statistical grounds.

Whether low-gravity T dwarfs exhibit stronger variability remains unknown, but \simp\ may provide the first such example with its $\sim$\,1--6\% $J$-band variability \citep{2016arXiv160903587C}. Only a handful of low-gravity T dwarfs are currently confirmed (e.g., \citealp{2015ApJ...808L..20G, 2014ApJ...787....5N}), none of which have reported measurements assessing their photometric variability.

\section{SUMMARY AND CONCLUSIONS}\label{sec:conclusion}

This paper presents the discovery that the nearby ($\sim$\,6\,pc) T2.5 dwarf \simp\ is a likely member of the 200\,Myr-old Carina-Near moving group, based on the new Bayesian analysis tool BANYAN~$\Sigma$ and radial velocity measurement. At this young age, \simp\ has a model-dependent mass of $12.7 \pm 1.0$\,\Mjup\ at the planetary-mass boundary. Given the tentative correlation between high-amplitude variability and youth in L dwarfs, the discovery that \simp\ is a member of Carina-Near indicates that such a correlation could hold in the T dwarfs regime, however more young T dwarfs will need to be investigated for variability to verify this. \simp\ is an even more powerful benchmark than previously appreciated and will help to understand weather patterns in gaseous giant atmospheres.

\begin{figure*}[!htbp]
\centering
\includegraphics[width=0.885\textwidth]{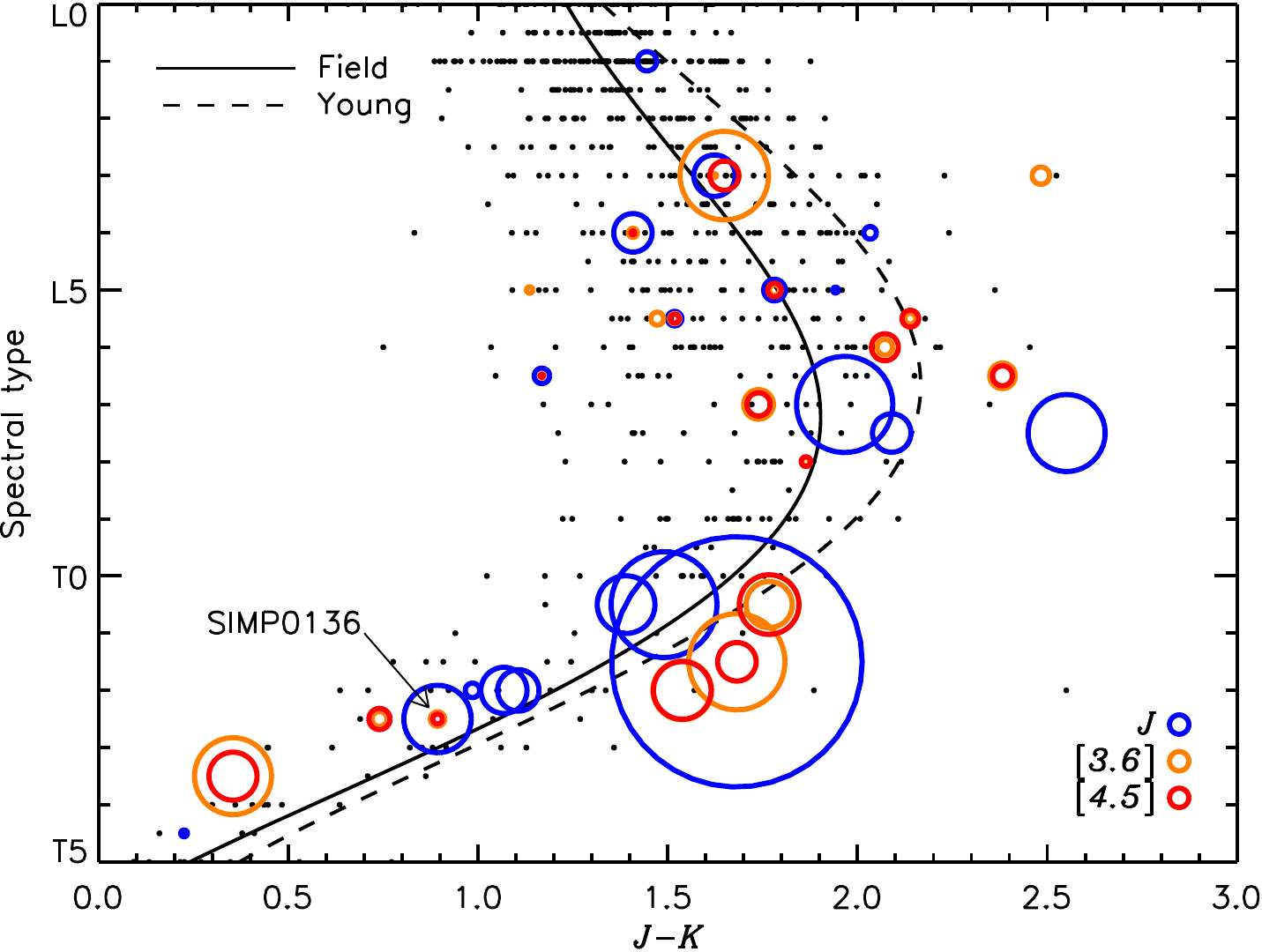}    
\caption{Brown dwarfs with detected photometric variability as a function of spectral type and near-infrared colors. The colors of individual circles indicate the wavelength where variability is detected, and the size of the circle indicates the variability amplitude in linear scale. The polynomial relations of \cite{2016ApJS..225...10F} for field and young brown dwarfs are displayed as solid and dashed lines, respectively. In the L spectral class, there is tentative evidence for high-amplitude variable brown dwarfs to be more frequent at redder near-infrared colors \citep{2015ApJ...799..154M}, which are generally associated with younger, low-gravity brown dwarfs. This figure was built with data from \cite{2012ApJS..201...19D}, \cite{2014ApJ...797..120R} and references therein.}
\label{fig:var}
\end{figure*}

\acknowledgments

The authors would like to thank the anonymous referee for useful comments and suggestions. This work was supported in part through grants from the Natural Science and Engineering Research Council of Canada. This research made use of: data products from the \emph{Wide-field Infrared Survey Explorer} (\emph{WISE}; \citealp{2010AJ....140.1868W}), which is a joint project of the University of California, Los Angeles, and the Jet Propulsion Laboratory at the California Institute of Technology (Caltech), funded by NASA, and of data from the European Space Agency (ESA) mission {\it Gaia} (\url{http://www.cosmos.esa.int/gaia}), processed by the {\it Gaia} Data Processing and Analysis Consortium (DPAC, \url{http://www.cosmos.esa.int/web/gaia/dpac/consortium}). The data presented herein were obtained at the W.M. Keck Observatory, which is operated as a scientific partnership among Caltech, the University of California and NASA. The Observatory was made possible by the generous financial support of the W.M. Keck Foundation. The authors wish to recognize and acknowledge the very significant cultural role and reverence that the summit of Mauna Kea has always had within the indigenous Hawaiian community. We are most fortunate to have the opportunity to conduct observations from this mountain.\\

\emph{JG} wrote most of the manuscript, generated Figures~\ref{fig:kinematics}, \ref{fig:sed} and \ref{fig:radius}, led the BASS-Ultracool survey and the development of the BANYAN~$\Sigma$ tool, and the kinematic analysis; \emph{JKF} led the spectral energy distribution analysis. \emph{AJB} acquired and reduced the NIRSPEC spectrum and measured the radial velocity and $v\sin i$. \emph{\'EA} wrote Section~\ref{sec:var} and generated Figures~\ref{fig:d2} and \ref{fig:var}. \emph{SB} provided useful discussions on the photometric properties of \simp\ and unpublished photometric data. \emph{LA} cross-matched \simp\ with individual epochs of the Pan-STARRS1 catalog. \emph{RD} and \emph{DL} participated in the development of the BANYAN~$\Sigma$ tool. \emph{DCBG} observed the NIRSPEC spectrum.

\facility{Keck:II (NIRSPEC)}
\software{BANYAN~$\Sigma$, BANYAN~II, Python, IDL by Harris Geospatial, Overleaf}

\bibliographystyle{apj}

\end{document}